\documentclass[preprint,12pt]{elsarticle}

\newcommand{\diffop}{\mathrm{d}}

\usepackage{lineno}

\usepackage{amsfonts}
\usepackage{amssymb}
\usepackage{xcolor}
\usepackage{amsmath}
\usepackage{booktabs}
\usepackage{siunitx}
\usepackage{subcaption}
\usepackage[section]{placeins}
\usepackage{hyperref}

\newcounter{bla}

\journal{Computer Physics Communications}

\begin{document}

\begin{frontmatter}

\title{GETaLM: A Generator for Electron Tagger and Luminosity Monitor for electron - proton and ion collisions}

\author{Jaroslav Adam}
\address{Brookhaven National Laboratory, Upton, United States}

\begin{abstract}
The study of elastic bremsstrahlung and electron tagging in electron-proton or ion collisions is gaining importance
with the planned construction of several experimental facilities focused on deep-inelastic scattering (DIS) measurements.
This paper describes a program
which generates
bremsstrahlung photons in electron-proton and electron-ion interactions as well as scattered electrons
in bremsstrahlung processes and in a quasi-real photon approximation to the general DIS process. The effects of electron
beam divergence
and the spread of the interaction vertex are implemented. The program can be used as an input to simulations of
instrumentation for bremsstrahlung photon detection, luminosity measurements, electron tagging,
and the determination of the cross sections of corresponding processes.
\end{abstract}


\end{frontmatter}

\let\thefootnote\relax\footnotetext{\textit{E-mail address:} jaroslavadam299@gmail.com}

{\bf PROGRAM SUMMARY}

\begin{small}
\noindent
{\em Program Title:} GETaLM \\
{\em Licensing provisions}: GNU GPLv3 \\
{\em Programming language:} Python \\
{\em External routines:} ROOT \\
{\em Nature of problem:} Photons due to relativistic bremsstrahlung processes are produced in collisions of electrons
with protons and with ions.
Detection of these bremsstrahlung photons is a promising method for luminosity measurements. The detection
of electrons scattered at small angles will impact these measurements.
The program generates the bremsstrahlung photons along with final state electrons in the bremsstrahlung
process and in an approximation to general electron-proton scattering.\\
{\em Solution method:} Analytic formulas for the cross sections of the specific processes and the relativistic kinematics
are used to generate the photons and scattered electrons. A set of effects imposed by the interacting beams can be applied
to the generated particles. The output of the program is created using the ROOT program.
Total cross sections are obtained by integrating the specific cross section formulas over a given kinematic region. \\
{\em Restrictions:} Currently the electron and proton (ion) beams are assumed to collide head-on, no crossing angle is considered. \\
{\em References:} \url{https://github.com/adamjaro/GETaLM} and references in this paper. \\
\end{small}

\section{Introduction}\label{sec:introduction}
The emission of bremsstrahlung photons in relativistic electron-proton and electron-ion collisions provide a convenient process
for the measurement of the luminosity. The photons are emitted at very small angles relative to the direction of the electron beam.
Several experimental facilities, the Electron-Ion Collider (EIC) \cite{AbdulKhalek:2021gbh},
the Large Hadron-Electron Collider \cite{Agostini:2020fmq} and the Electron-Ion Collider in China \cite{Anderle:2021wcy}
plan to have dedicated instrumentation to measure luminosity by detecting the bremsstrahlung photons and to tag electrons
scattered at small angles. The detection of scattered electrons is also used to constrain the kinematics of DIS processes.

The Generator for the Electron Tagger and Luminosity Monitor (GETaLM), presented here, is based on a set of analytic 
formulas describing the process dynamics and on relativistic kinematics to create the final bremsstrahlung photons 
and scattered electrons. A similar simulation package was outlined for the HERA collider \cite{Levonian:1993}.
The generator is configured
from an INI file and the output contains variables related to a particular process and a complete set of kinematic variables
for the final photons and electrons. The implementation is done using the ROOT program \cite{Brun:1997pa}. A prototype
of GETaLM was already used in the EIC study in Ref.~\cite{AbdulKhalek:2021gbh}.

\section{Elastic bremsstrahlung}\label{sec:brem}
\begin{figure}[!ht]
  \centering
  \includegraphics[width=0.8\textwidth]{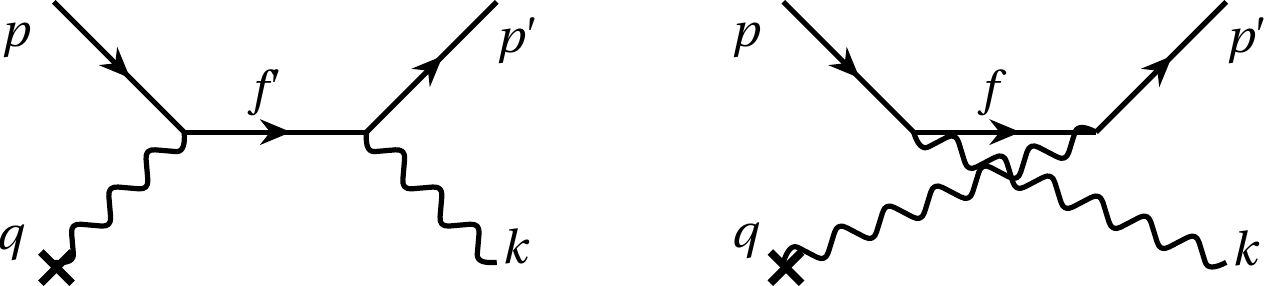}
  \caption{The lowest non-vanishing diagrams for the bremsstrahlung process.}
  \label{fig:diag-bh}
\end{figure}

The first non-vanishing Feynman diagrams for elastic bremsstrahlung are shown in Fig.~\ref{fig:diag-bh}.
An incoming electron (positron) of momentum $p$
scatters off the external field of the target particle (proton or nucleus), leaving an electron with a momentum $p'$
and a bremsstrahlung photon
of momentum $k$ in the final state.

Momentum transfer to the target particle is $q=p'-p+k$, the momenta of intermediate states are $f=p-k$ and $f'=p'+k$.

Several approximations to the process in Fig.~\ref{fig:diag-bh} which neglect the recoil of the target
particle ($q^0=0$) are given in the following sections.

\subsection{Parameterizations of the bremsstrahlung cross section}
The cross section for electron and proton beams of energy $E_e$ and $E_p$ is given by two equivalent parameterizations
in Eq.~\ref{eq:brem-zeus} as a function of bremsstrahlung photon energy $E_\gamma$ \cite{Haas:2010bq} and in Eq.~\ref{eq:brem-h1}
as a function of $y=E_\gamma/E_e$ \cite{Levonian:1993}.

\begin{equation}\label{eq:brem-zeus}
  \frac{\diffop\sigma}{\diffop E_\gamma} = 4\alpha r_e^2\frac{E'_e}{E_\gamma E_e}\left (\frac{E_e}{E'_e}+\frac{E'_e}{E_e}-\frac{2}{3}\right )
  \left (\ln\frac{4E_pE_eE'_e}{m_pm_eE_\gamma} - \frac{1}{2} \right )
\end{equation}

\begin{equation}\label{eq:brem-h1}
  \frac{\diffop\sigma}{\diffop y} = \frac{4\alpha r_e^2}{y}\left [ 1+(1-y)^2 - \frac{2}{3}(1-y) \right ]
  \left [ \ln\frac{s(1-y)}{m_p m_e y} - \frac{1}{2} \right ]
\end{equation}

The center-of-mass energy squared is $s$, the electron and proton rest masses are $m_e$ and $m_p$. Normalization of the cross
section is given by $4\alpha r_e^2$ = 2.3179~mb where $\alpha$ is the fine structure constant and $r_e$ is the classical electron radius.

The angular distribution of bremsstrahlung photons is given by Eq.~\ref{eq:brem-theta}. The angle $\theta_\gamma$ is the angle
of the 3-momentum component of the outgoing photon $k$ relative to the 3-momentum component of the incoming electron $p$.

\begin{equation}\label{eq:brem-theta}
  \frac{\diffop\sigma}{\diffop\theta_\gamma}\sim\frac{\theta_\gamma}{\left ((m_e/E_e)^2 + \theta_\gamma^2 \right )^2}
\end{equation}

\begin{figure}
  \centering
  \includegraphics[width=0.6\textwidth]{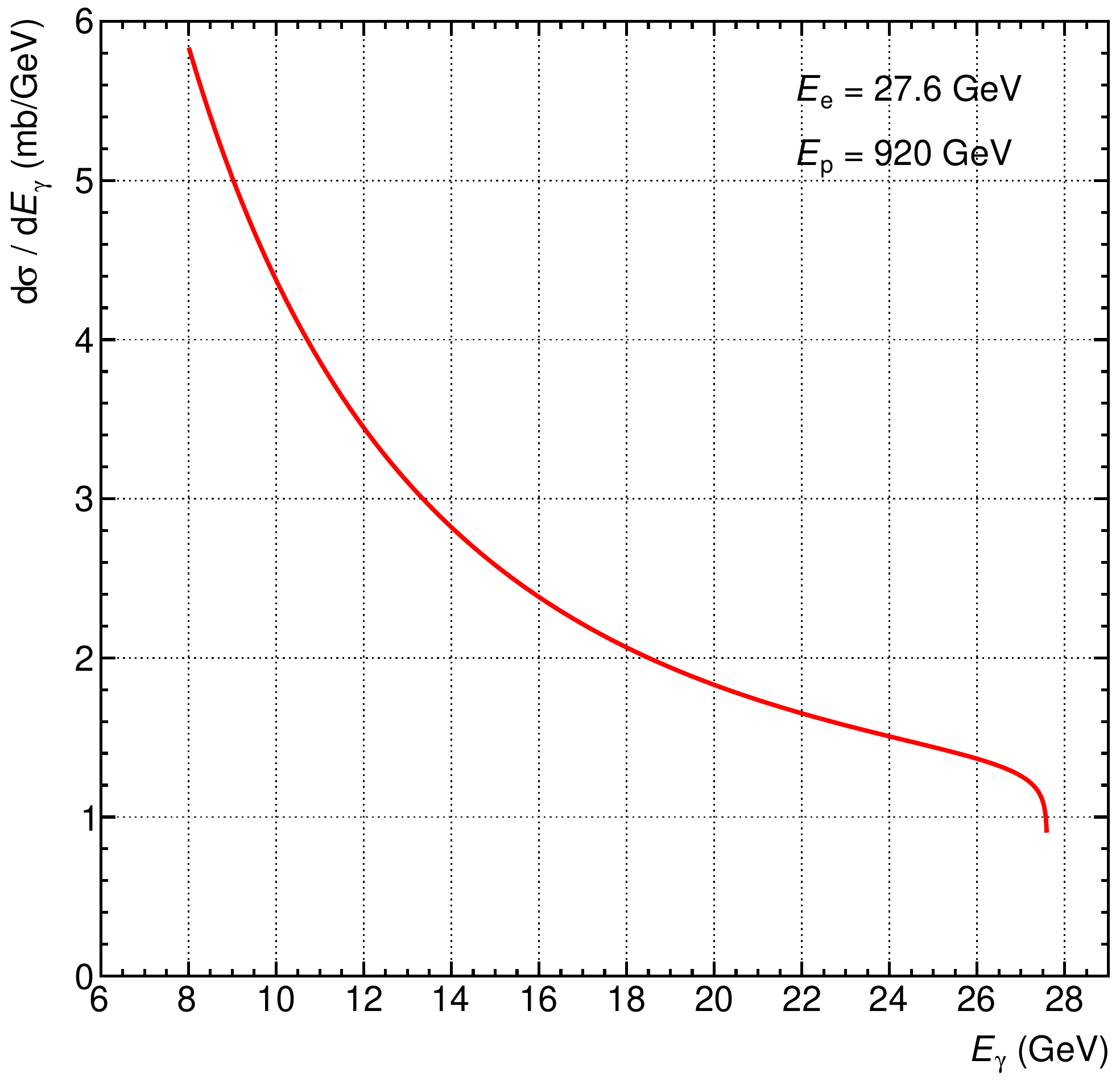}
  \caption{Bremsstrahlung cross section as a function of energy given by Eq.~\ref{eq:brem-zeus}.}
  \label{fig:sigma-Eg}
\end{figure}

The cross section $\diffop \sigma/\diffop E_\gamma$ according to Eq.~\ref{eq:brem-zeus} is shown in Fig.~\ref{fig:sigma-Eg}
for $E_e$ = 27.6~GeV and $E_p$ = 920~GeV, an energy corresponding to that of the HERA collider. This result is compatible with a similar
calculation in Ref.~\cite{Adamczyk:2013ewk}.

\subsection{QED calculation in electron-nucleus case}
The doubly-differential bremsstrahlung cross section for the general case of a nucleus
of electric charge $Z$ is given in Eq.~\ref{eq:brem-bh} \cite{Berestetskii:1982}.

\begin{align}\label{eq:brem-bh}
  \frac{\mathrm{d}^2\sigma}{\mathrm{d}\omega\mathrm{d}\delta} &= 8Z^2\alpha r_e^2\frac{1}{\omega}\frac{\varepsilon'}{\varepsilon}
  \frac{\delta}{(1+\delta^2)^2} \times \nonumber \\
  &\times \left\{ \left[ \frac{\varepsilon}{\varepsilon'} + \frac{\varepsilon'}{\varepsilon}
  -\frac{4\delta^2}{(1+\delta^2)^2} \right] \ln\frac{2\varepsilon\varepsilon'}{m_e\omega}
  -\frac{1}{2} \left[ \frac{\varepsilon}{\varepsilon'} + \frac{\varepsilon'}{\varepsilon} +2
  -\frac{16\delta^2}{(1+\delta^2)^2} \right] \right\}
\end{align}

The energy of the bremsstrahlung photon is $\omega$ and the electron initial and final energies are $\varepsilon$
and $\varepsilon'$ respectively,
all given in the rest frame of the target nucleus. The angle of the photon $\vartheta_\gamma$ relative to the initial electron,
also in the target nucleus rest frame, is obtained from $\delta=\varepsilon\vartheta_\gamma/m_e$.

The integrated bremsstrahlung cross section for $E_\gamma >$ 0.1~GeV $\sigma_{\mathrm{brem}}$ obtained
from Eq.~\ref{eq:brem-bh} is shown
in Table~\ref{tab:sig-lif}.
The cross sections are evaluated for a set of electron-proton ($ep$) and electron-gold nucleus ($e$-Au) energies
considered for the EIC.

\begin{table}[h]
\begin{center}
\begin{tabular}{l c c|c c|c c|c c}
\toprule
Species & $e$ & $p$ & $e$ & $p$ & $e$ & Au & $e$ & Au\\
Energy (GeV) & 18 & 275 & 5 & 41 & 18 & 110 & 5 & 41\\
\midrule
$\sigma_{\mathrm{brem}}$ & \multicolumn{2}{c|}{276.3 mb} & \multicolumn{2}{c|}{159.3 mb} & \multicolumn{2}{c|}{1.563 kb}
& \multicolumn{2}{c}{0.936 kb}\\
\bottomrule
\end{tabular}
\end{center}
\caption{Integrated bremsstrahlung cross section $\sigma_{\mathrm{brem}}$ predicted by Eq.~\ref{eq:brem-bh} for a set
of $ep$ and $e$-Au energies.}
\label{tab:sig-lif}
\end{table}

The different units of mb and kb for $\sigma_{\mathrm{brem}}$ should be noted in Table~\ref{tab:sig-lif} reflecting very large
bremsstrahlung cross sections in the case of ion beams.

\section{Quasi-real photoproduction}\label{sec:qr}
\begin{figure}[!ht]
  \centering
  \includegraphics[width=0.35\textwidth]{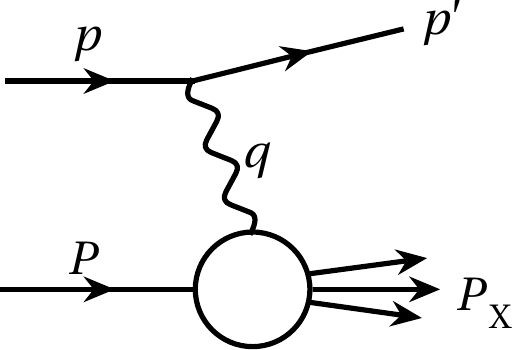}
  \caption{Deep inelastic scattering.}
  \label{fig:diag-dis}
\end{figure}

The deep inelastic scattering of a lepton of momentum $p$ off a nucleon of momentum $P$ to a final lepton $p'$ and a hadronic system
of momentum $P_X$ is shown in Fig.~\ref{fig:diag-dis}.

The DIS process is described by three independent kinematic variables. The center-of-mass energy squared is $s=(p+P)^2$, virtuality
of the exchanged boson is $q^2 = -Q^2 = (p-p')^2$ and the center-of-mass energy of the boson-nucleon system is $W^2=(P+q)^2$.

The Bjorken $x$ is $x=Q^2/2Pq$ and inelasticity is $y=Pq/Pp$ giving the fraction of initial lepton energy carried
by the exchanged boson.
The kinematic variables are related by

\begin{equation}\label{eq:Q2xys}
  Q^2 = xys
\end{equation}
when the rest masses are neglected.

In the approximation to quasi-real photoproduction off protons, where the virtuality $Q^2$ of the exchanged photon
is negligible, the cross section
for the process in Fig.~\ref{fig:diag-dis} is

\begin{equation}\label{eq:qr-II6}
  \frac{\diffop^2\sigma}{\diffop x \diffop y} = \frac{\alpha}{2\pi}\ \frac{1+(1-y)^2}{y}
  \ \sigma_{\gamma p}(W^2)\ \frac{1-x}{x}
\end{equation}
where $\sigma_{\gamma p}$ is the total photon-proton cross section \cite{Amaldi:1979qp, Frixione:1993yw}.

The experimental photon-proton cross section is \cite{Donnachie:1992ny}

\begin{equation}\label{eq:qr-sigmaGp}
  \sigma_{\gamma p}(W^2) = 0.0677\left( W^2 \right)^{0.0808} + 0.129\left( W^2 \right)^{-0.4525}\ (\mathrm{mb}).
\end{equation}

Table~\ref{tab:sig-qrpy} gives a comparison of the integrated cross section $\sigma_\mathrm{tot}$ between the quasi-real approximation
in Eq.~\ref{eq:qr-II6} with $\sigma_{\gamma p}$ from Eq.~\ref{eq:qr-sigmaGp} and the PYTHIA~6 \cite{Sjostrand:2006za} event generator.
The comparison is made for a set of electron $E_e$ and proton $E_p$ beam energies considered for the EIC.

\begin{table}[h]
\begin{center}
\begin{tabular}{c c c c}
\toprule
\multicolumn{2}{c}{Energy (GeV)} & \multicolumn{2}{c}{$\sigma_\mathrm{tot}$ (\si{\micro\barn})}\\
$E_e$ & $E_p$ & Quasi-real & PYTHIA 6\\
\midrule
18 & 275 & 55.1 & 54.7\\
10 & 100 & 44.8 & 40.9\\
5 & 41 & 33.4 & 28.4\\
\bottomrule
\end{tabular}
\end{center}
\caption{Integrated cross section of quasi-real photoproduction and comparison to PYTHIA~6.}
\label{tab:sig-qrpy}
\end{table}

The integration range for both the quasi-real and the PYTHIA~6 results
is $10^{-11} < x < 1$, $10^{-4} < y < 0.99$, $Q^2>$ $10^{-9}$~GeV$^2$
and $W>$ 2~GeV.
The actual lower limit on $y$ is imposed by the lower limit on $W$ following the relation $W^2=ys$ which holds
at low $Q^2$ and when the rest masses are neglected.

\section{Particle generation}\label{sec:particle-gen}

The formalism outlined in Sec.~\ref{sec:brem} and Sec.~\ref{sec:qr} is used to generate the momenta of bremsstrahlung photons,
electrons
scattered in the bremsstrahlung process and electrons in the final state of quasi-real photoproduction.

\subsection{Bremsstrahlung photons from parameterizations}
The cross section formulas in Eq.~\ref{eq:brem-zeus} and Eq.~\ref{eq:brem-h1} are used to obtain the values of bremsstrahlung
photon energy $E_\gamma$. The energy of the final state electron $E'_e$ in Eq.~\ref{eq:brem-zeus} is set as $E'_e=E_e-E_\gamma$.
In the case of Eq.~\ref{eq:brem-h1}, the relation $E_\gamma=yE_e$ is used.

In a similar way, the values of the photon polar angle $\theta_\gamma$ are generated from Eq.~\ref{eq:brem-theta}. Azimuthal angles
$\phi_\gamma$ are generated as uniform over the full range $0<\phi_\gamma<2\pi$.

With a set of values for $E_\gamma$, $\theta_\gamma$ and $\phi_\gamma$ it is possible to set the photon momentum $k$ in the final state
shown in Fig.~\ref{fig:diag-bh}. The final electron momentum is then approximated as $p'=p-k$.

\subsection{Bremsstrahlung photons in QED for the electron-nucleus case}\label{sec:gen-brem}
Equation~\ref{eq:brem-bh} is used to simultaneously generate values of photon energy $\omega$ and polar angle $\vartheta_\gamma$
given in the rest frame of target nucleus. The relation $\delta=\varepsilon\vartheta_\gamma/m_e$ is used for the polar angle.

The photon azimuthal angles in the nucleus rest frame are generated uniformly over the full range, $0<\varphi_\gamma<2\pi$.

The energy $\omega$ and angles $\vartheta_\gamma$ and $\varphi_\gamma$ are used to set the photon momentum in the nucleus rest frame,
which is then boosted to the laboratory frame to obtain the final photon momentum $k$. As in the previous case the final electron
momentum is approximated as $p'=p-k$.

\begin{figure}
  \centering
  \includegraphics[width=0.6\textwidth]{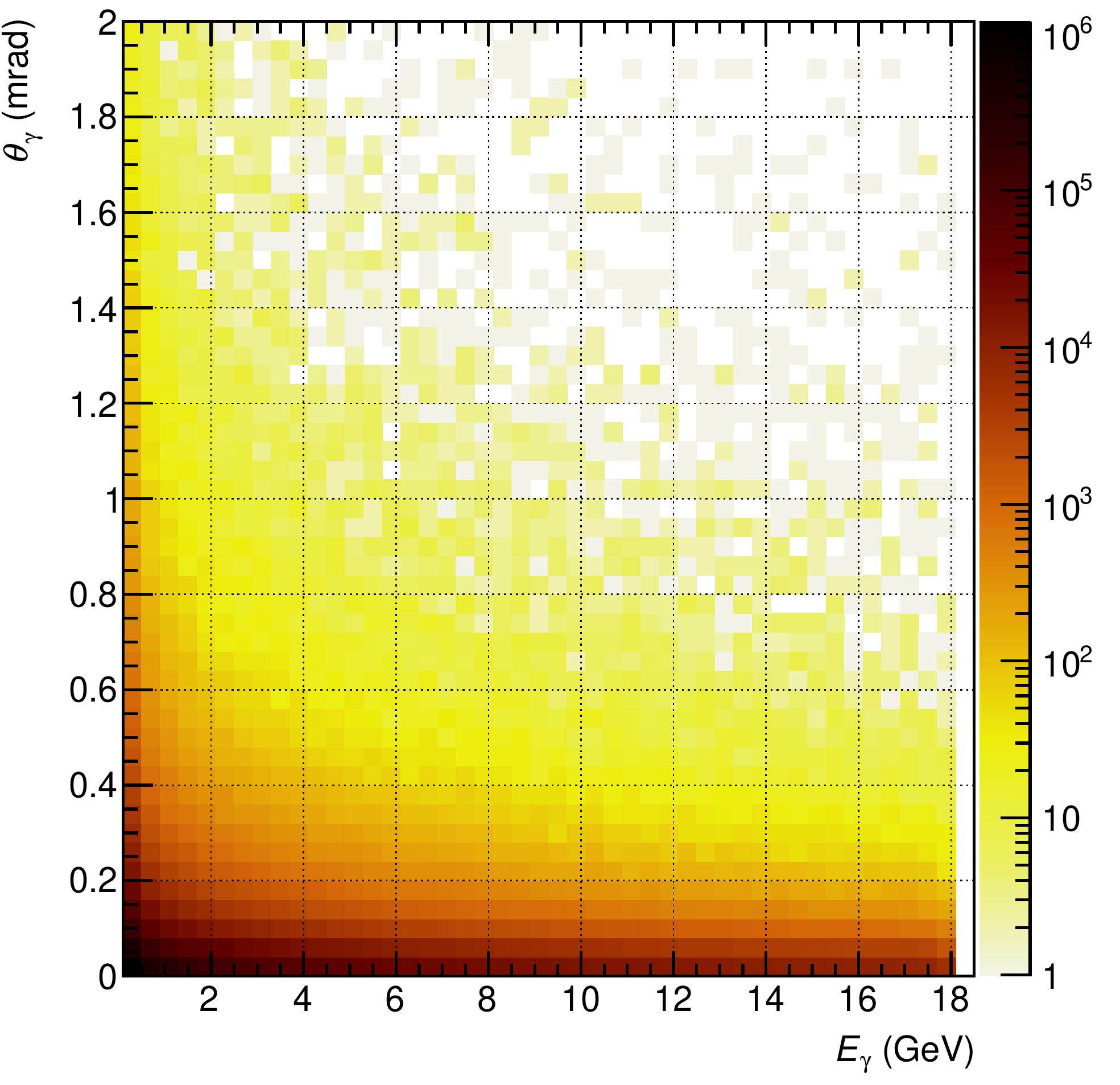}
  \caption{Energy $E_\gamma$ and polar scattering angle $\theta_\gamma$ of the bremsstrahlung photons for beam energies
  $E_e$ = 18~GeV and $E_p$ = 275~GeV.}
  \label{fig:en-theta}
\end{figure}

The photon polar angle $\theta_\gamma$ corresponding to a momentum $k$ is no longer independent of its energy $E_\gamma$, given
the nature of double-differential cross section in Eq.~\ref{eq:brem-bh}. The distribution of the generated photon angles $\theta_\gamma$
and energies $E_\gamma$ is shown in Fig.~\ref{fig:en-theta}.

\subsection{Electrons in quasi-real photoproduction}
The cross section for quasi-real photoproduction in Eq.~\ref{eq:qr-II6} is transformed to decadic logarithms of Bjorken $x$
and inelasticity $y$ as

\begin{equation}
\begin{aligned}\label{eq:qr-transform}
  u &= \log_{10}(x)\\
  v &= \log_{10}(y)
\end{aligned}
\end{equation}
to increase the precision of generation at low values of $x$ and $y$.

The values of $W^2$ at which the photon-proton cross section $\sigma_{\gamma p}$ is evaluated with Eq.~\ref{eq:qr-sigmaGp}
are set as $W^2 = ys$, or $W^2 = 10^vs$ after the transformation.

After the transformation in Eq.~\ref{eq:qr-transform} the cross section Eq.~\ref{eq:qr-II6} becomes

\begin{equation}\label{eq:qr-II6-uv}
  \frac{\diffop^2\sigma}{\diffop u \diffop v} = \frac{\alpha}{2\pi}\left[ 1+(1-10^v)^2 \right]\cdot
  \sigma_{\gamma p}(10^vs)\cdot (1-10^u).
\end{equation}

Values for $x$ and $y$ are generated from Eq.~\ref{eq:qr-II6-uv} and the inverse of Eqs.~\ref{eq:qr-transform}, then the momentum
of the scattered electron is found in the proton rest frame and boosted to the laboratory frame.

To find the scattered electron energy $\varepsilon'$ and angle $\vartheta_e$ in the proton rest frame the kinematic relations
\begin{align}
y &= (\varepsilon - \varepsilon')/\varepsilon \label{eq:qr-kine-y}\\
Q^2 &= 4\varepsilon \varepsilon' \sin^2\frac{\vartheta_e}{2} \label{eq:qr-kine-Q2}
\end{align}
are used. $\varepsilon$ is the initial electron energy in the proton rest frame.

With Eqs.~\ref{eq:Q2xys} and \ref{eq:qr-kine-y} $\vartheta_e$ can be expressed
as a function of $x$ and $y$ alone as

\begin{equation}\label{eq:qr-theta-xy}
  \vartheta_e = 2\arcsin\left( \frac{1}{2}\sqrt{\frac{xys}{(1-y)\varepsilon^2}} \right).
\end{equation}

Using the generated values for $x$ and $y$, the electron energy $\varepsilon'$ and angle $\vartheta_e$ are obtained from
Eq.~\ref{eq:qr-kine-y} and Eq.~\ref{eq:qr-theta-xy} respectively. The electron azimuthal angle in the proton rest frame $\varphi_e$
is generated uniformly over the full range $0<\varphi_e<2\pi$.

The electron momentum in the proton rest frame is constructed from $\varepsilon'$, $\vartheta_e$, $\varphi_e$ and the electron rest mass
$m_e$. Then it is boosted to the laboratory frame to get the final electron momentum $p'$ in Fig.~\ref{fig:diag-dis}.

\begin{figure}
  \centering
  \begin{subfigure}[h]{0.49\textwidth}
    \includegraphics[width=0.9\textwidth]{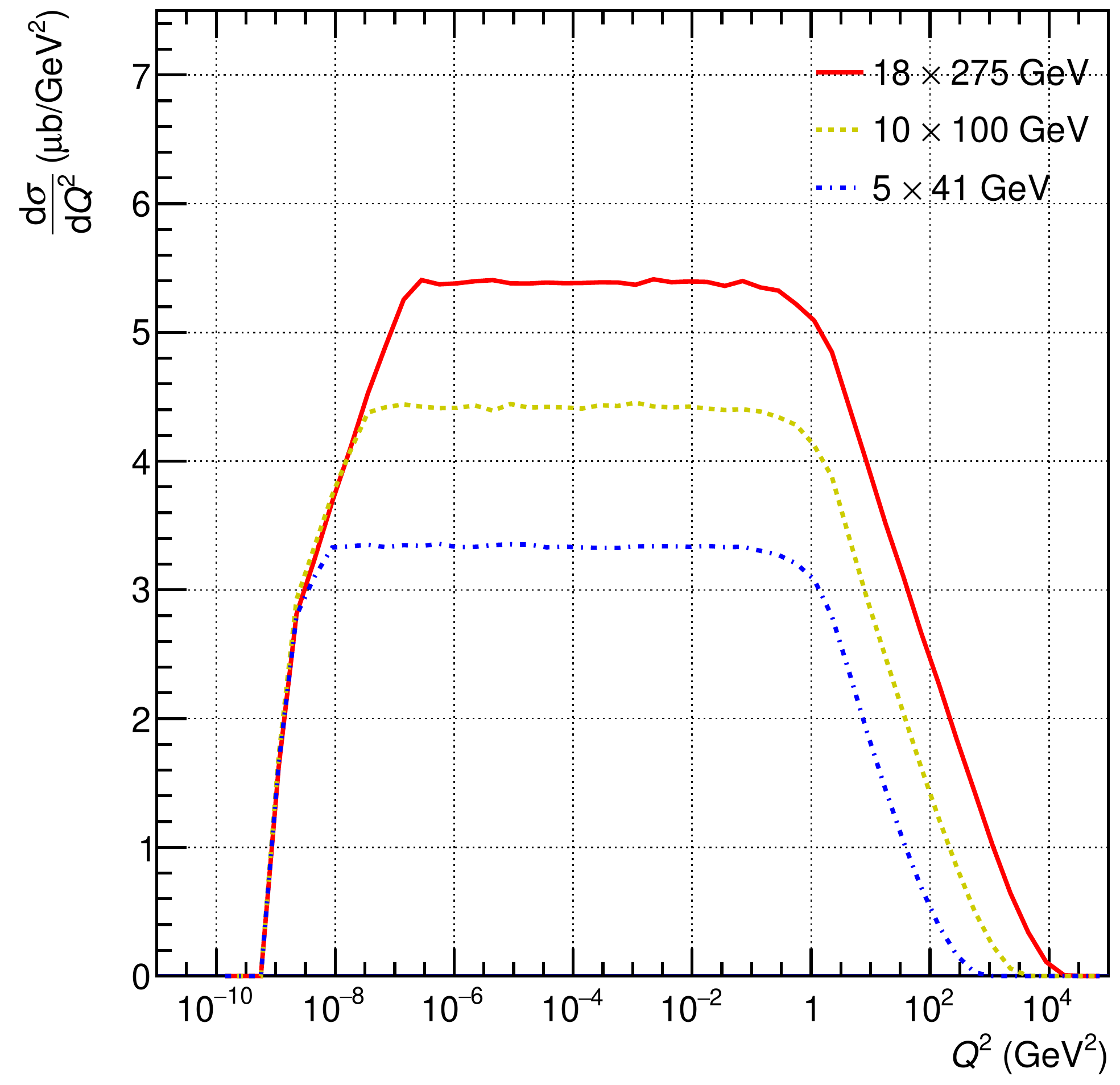}
    \caption{}
    \label{fig:Q2-sigma}
  \end{subfigure}
  \begin{subfigure}[h]{0.49\textwidth}
    \includegraphics[width=0.9\textwidth]{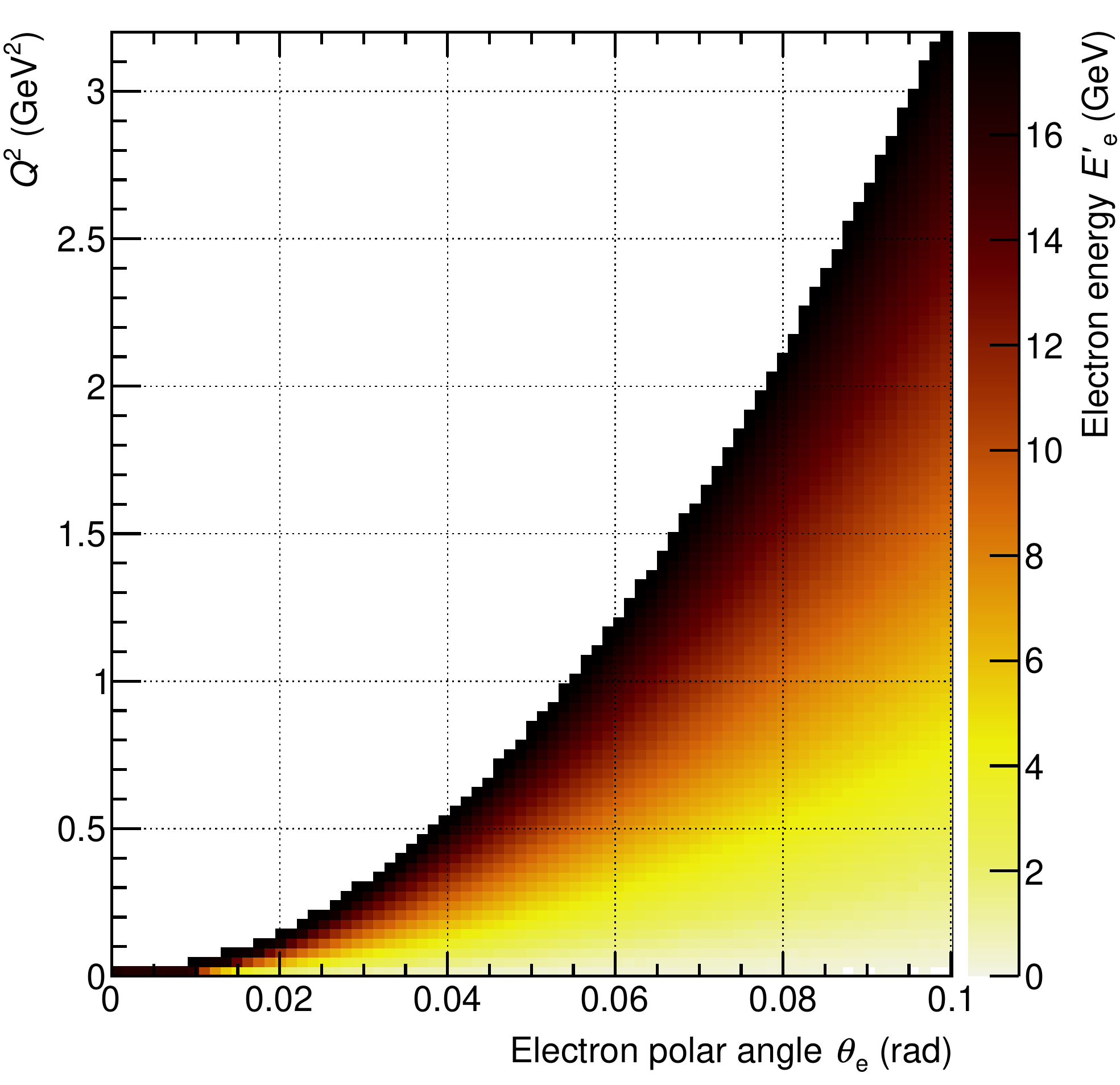}
    \caption{}
    \label{fig:Q2-theta-en}
  \end{subfigure}
  \caption{Total cross section of quasi-real photoproduction as a function of $Q^2$ for several beam energies (a)
  and the relation between $Q^2$, $\theta_e$ and $E'_e$ for beam energies $E_e$ = 18~GeV and $E_p$ = 275~GeV (b).}
  \label{fig:Q2-sigma-theta-en}
\end{figure}

The cross section of quasi-real photoproduction corresponding to Table~\ref{tab:sig-qrpy} is shown in Fig.~\ref{fig:Q2-sigma}
as a function of $Q^2$, where the $Q^2$ is determined by Eq.~\ref{eq:Q2xys}.

The relation between $Q^2$ and the final electron energy $E'_e$ and polar angle $\theta_e$ is shown in Fig.~\ref{fig:Q2-theta-en}
with the energy given by the color scale. The initial kinematics for this figure corresponds
to the first line of Table~\ref{tab:sig-qrpy}.

\subsection{Effects of vertex spread and beam angular divergence}\label{sec:beff}
In colliding beams the position of interaction (the vertex) depends on the actual profile of interacting beams.
The result is a spread in the vertex position both along the beams and in the transverse directions.

The generator can create vertex positions following three separate Gaussian distributions for the direction along the beams
and in the transverse directions. Each of the distributions are centered at the origin and are described by a specific
width.

The angular divergence of the beam amounts to the spread in the initial angles of particles
in the beam interacting with the particles
of the opposite beam.

The generator implements the divergence of the electron beam by perturbing the 3-momenta of the final particles. Random
Gaussian rotations are imposed on the particles' 3-momenta with a width $\sigma_{\mathrm{div}}$ in the $x$-direction
and another in the $y$-direction.

\begin{figure}
  \centering
  \includegraphics[width=0.6\textwidth]{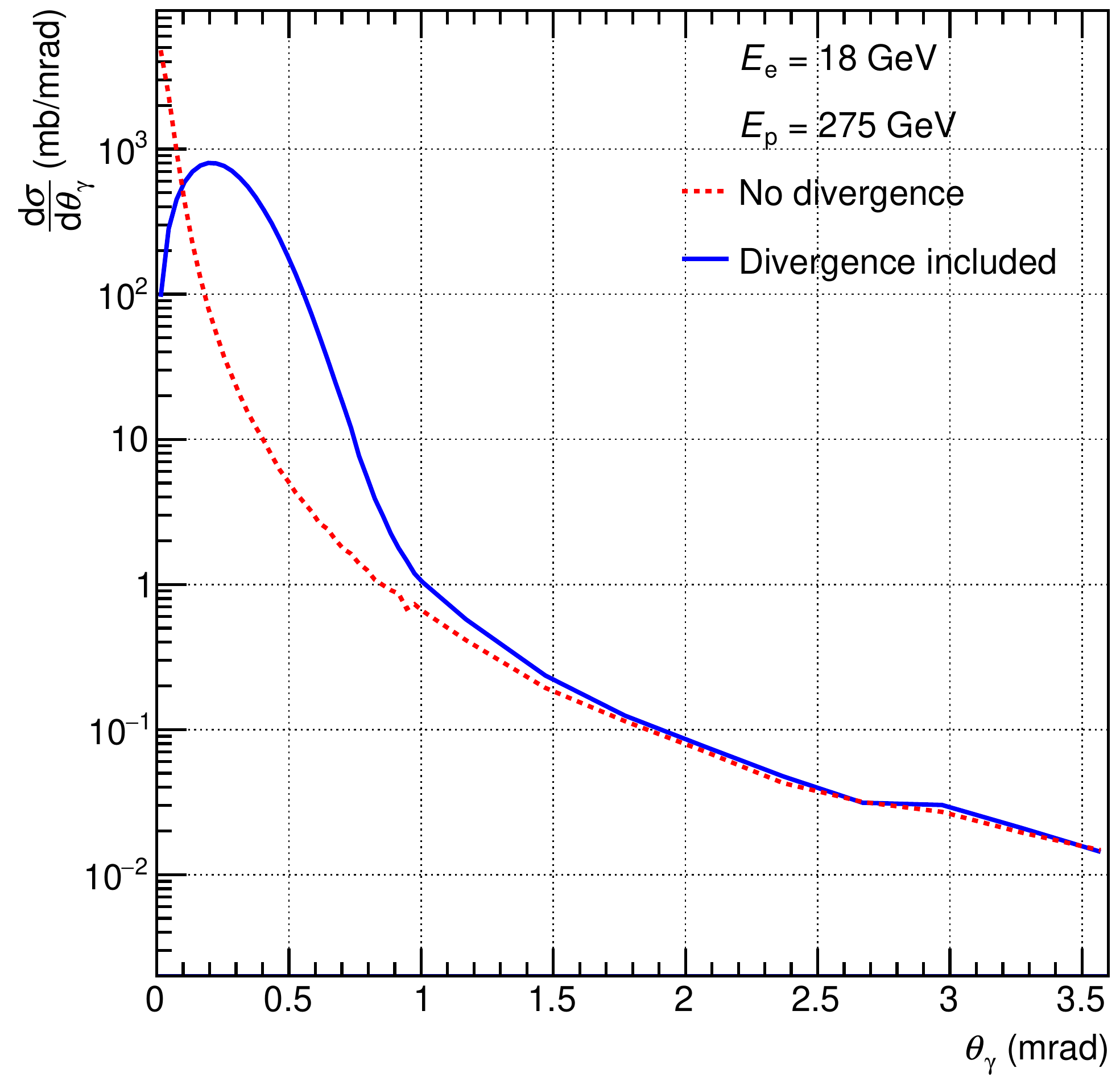}
  \caption{Bremsstrahlung cross section as a function of photon polar angle $\theta_\gamma$ with an illustration of the effect
  of beam angular divergence.}
  \label{fig:theta-div}
\end{figure}

The effect of the electron beam divergence on the polar angles $\theta_\gamma$ of bremsstrahlung photons (Sec.~\ref{sec:gen-brem})
is illustrated in Fig.~\ref{fig:theta-div}. The bremsstrahlung cross section as a function of photon angle
$\diffop\sigma/\diffop\theta_\gamma$ is shown for the case when the divergence is not considered and for the case
of the divergence of $\sigma_{\mathrm{div}}$ = 202~\si{\micro\radian} in both transverse directions is included.

\begin{figure}
  \centering
  \includegraphics[width=0.6\textwidth]{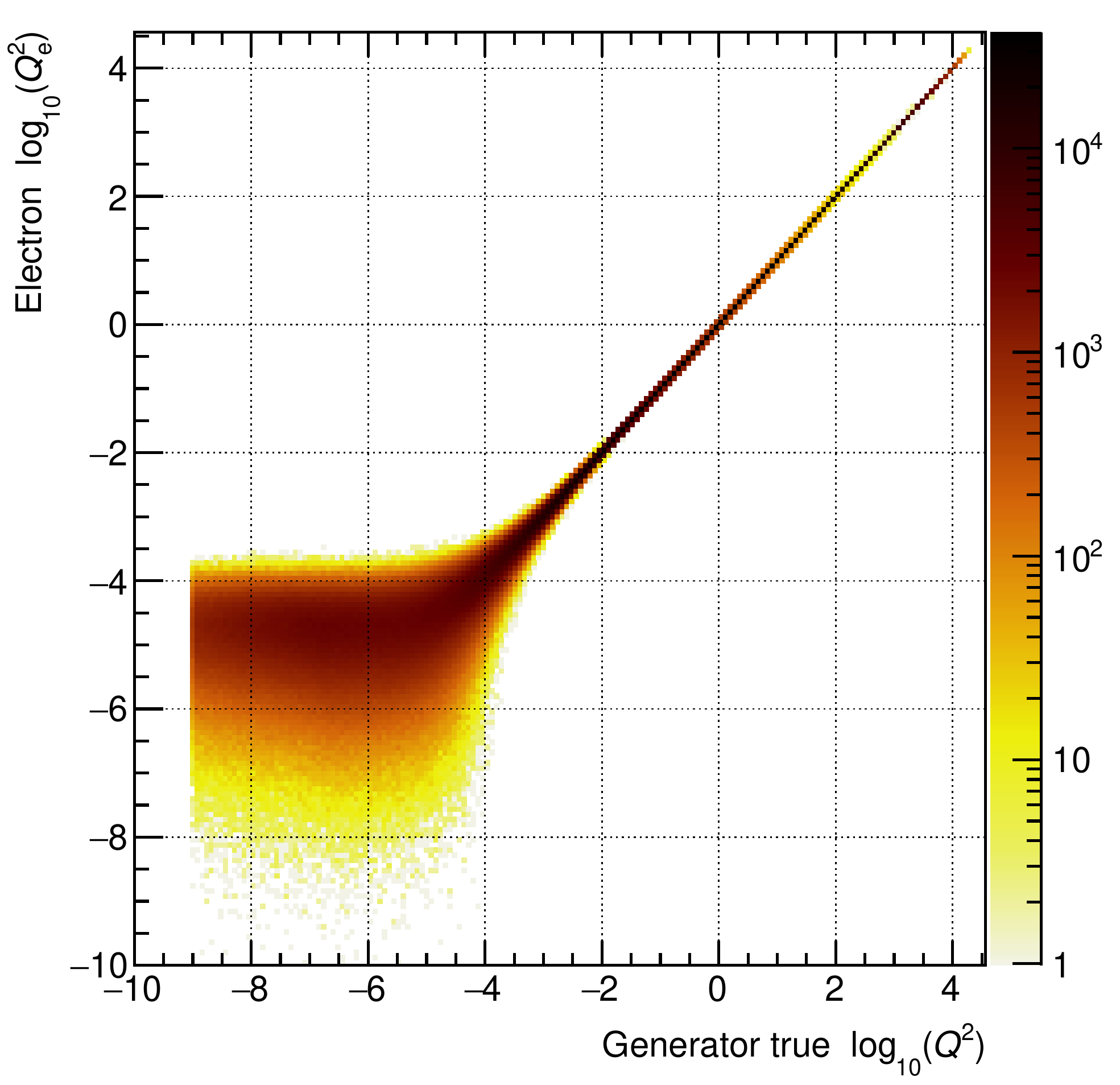}
  \caption{Comparison of $Q^2$ generated in the event (true $Q^2$) and $Q^2_e$ given by the electron kinematics affected by the beam
  angular divergence.}
  \label{fig:trueQ2-elQ2}
\end{figure}

The electron beam divergence has a strong effect at low values of $Q^2$, as illustrated in Fig.~\ref{fig:trueQ2-elQ2}
for a sample of events of quasi-real photoproduction with $\sigma_{\mathrm{div}}$ = 202~\si{\micro\radian} in both transverse
directions.
The virtuality $Q^2_e$ determined from the scattered electron
as $Q^2_e = 2E_eE'_E(1-\cos(\theta_e))$ including the divergence is compared to the true generated $Q^2$ given by
Eq.~\ref{eq:Q2xys}.
The correspondence between $Q^2_e$ and $Q^2$ is nearly perfect for $Q^2 \gtrsim$ $10^{-2}$~GeV$^2$ but is completely lost
at $Q^2 \lesssim$ $10^{-4}$~GeV$^2$.

\section{Program flow}
The actions carried out by the generator program are described in this section.
\begin{itemize}
  \item Initialization
  \begin{itemize}
    \item Configuration is loaded from the input steering card, provided by the user.
    \item One of the physics models in Sec.~\ref{sec:brem} and Sec.~\ref{sec:qr} is loaded and initialized.
  \end{itemize}
  \item Event generation
  \begin{itemize}
    \item A model-specific method for particle generation is called for the loaded physics model.
    \item The model generates the particles according to Sec.~\ref{sec:particle-gen}.
    \item Optionally, effects of vertex spread and beam angular divergence are applied to generated particles
          according to Sec~\ref{sec:beff}
  \end{itemize}
  \item Generator output
  \begin{itemize}
    \item Event-wide variables, specific to the models are written as numerical values to the event \texttt{TTree}.
    \item Particles are written as a \texttt{TClonesArray} of \texttt{TParticle} objects in the event \texttt{TTree}.
          The coordinate convention of \cite{AbdulKhalek:2021gbh} is used for the particles, namely that the positive $z$ direction
          points in the direction of the proton (ion) beam.
  \end{itemize}
\end{itemize}

\section{Description of steering card}
The generator is configured from a steering card in the INI file. The file is parsed
by the \texttt{ConfigParser} module. Parameters related to physics models and generator output are grouped
in the \texttt{[main]} section, parameters describing effects of vertex spread and beam angular divergence
belong to the \texttt{[beam\_effects]} section. The \texttt{[beam\_effects]} section is optional.
Default values for the parameters are indicated in parenthesis.

\begin{itemize}
  \item General run options
  \begin{itemize}
    \item \texttt{Ee}, \texttt{Ep}: Energy of electron and proton (ion) beam respectively in GeV.
    \item \texttt{nev}: Number of events to generate.
    \item \texttt{nam}: Name for the output file. The .root extension is appended automatically.
    \item \texttt{model}: Physics model for particle generation.
  \end{itemize}
  \item Options for bremsstrahlung models
  \begin{itemize}
    \item The models are selected by the \texttt{model} parameter set to 'zeus' for particle generation based on Eq.~\ref{eq:brem-zeus},
          'h1' for Eq.~\ref{eq:brem-h1} or 'Lifshitz\_93p16' for Eq.~\ref{eq:brem-bh}
    \item \texttt{emin}: Minimum bremsstrahlung photon energy $E_\gamma$ in GeV
    \item \texttt{tmax}: Maximum photon polar angle $\theta_\gamma$ in rad, available only for 'zeus' model ($1.5\times 10^{-3}$~rad).
    \item \texttt{A}, \texttt{Z}: Target nucleus, available only for 'Lifshitz\_93p16' model (A = 1 and Z = 1).
    \item \texttt{dmax\_n}: Maximum value for $\delta$ in Eq.~\ref{eq:brem-bh}, the value determines maximum polar
          angle $\theta_\gamma$. Available only for 'Lifshitz\_93p16' model (100).
  \end{itemize}
  \item Options for quasi-real photoproduction
  \begin{itemize}
    \item The process of quasi-real photoproduction is selected by the \texttt{model} parameter set to 'quasi-real'.
    \item \texttt{xmin}, \texttt{xmax}: Minimum and maximum values for the Bjorken $x$.
    \item \texttt{ymin}, \texttt{ymax}: Minimum and maximum values for the inelasticity $y$.
    \item \texttt{Q2min}, \texttt{Q2max}: Minimum and maximum values for $Q^2$ in GeV$^2$.
    \item \texttt{Wmin}: Minimum value for $W$ in GeV, ignored when negative value is set (-1).
    \item \texttt{Wmax}: Maximum value for $W$ in GeV, ignored when negative value is set (-1).
  \end{itemize}
  \item Options for the effects of vertex spread and beam angular divergence
  \begin{itemize}
    \item \texttt{sig\_x}, \texttt{sig\_y}, \texttt{sig\_z}: Width of vertex spread along the $x$, $y$ and $z$ coordinates respectively.
    \item \texttt{theta\_x}, \texttt{theta\_y}: Width of angular divergence in the $x$ and $y$ coordinates in rad.
    \item \texttt{use\_beam\_effects}: Value of \texttt{true} to apply the effects, \texttt{false} otherwise (\texttt{false}).
  \end{itemize}
\end{itemize}

\section{Description of test data}
The parameters listed in Table~\ref{tab:card-brem} were used for Fig.~\ref{fig:theta-div}, the parameters in Table~\ref{tab:card-qr}
were used to produce Fig.~\ref{fig:Q2-sigma-theta-en}.
\begin{table}[h]
\begin{tabular}{l}
\toprule
$[\mathrm{main}]$\\
Ee = 18\\
Ep = 275\\
nev = 5000000\\
nam = "luminosity"\\
model = "Lifshitz\_93p16"\\
emin = 0.1\\
dmax\_n = 200\\
$\mathrm{[beam\_effects]}$\\
use\_beam\_effects = true\\
sig\_x = 0.119\\
sig\_y = 0.01\\
sig\_z = 9\\
theta\_x = 202e-6\\
theta\_y = 202e-6\\
\bottomrule
\end{tabular}
\caption{Example steering card for bremsstrahlung photons generation.}
\label{tab:card-brem}
\end{table}

\begin{table}[h]
\begin{tabular}{l}
\toprule
$[\mathrm{main}]$\\
Ee = 18\\
Ep = 275\\
nev = 5000000\\
nam = "dis"\\
model = "quasi-real"\\
xmin = 1e-11\\
xmax = 1\\
ymin = 1e-4\\
ymax = 0.99\\
Q2min = 1e-9\\
Q2max = 1e7\\
Wmin = 2\\
\bottomrule
\end{tabular}
\caption{Example steering card for quasi-real photoproduction.}
\label{tab:card-qr}
\end{table}
\FloatBarrier

\section{Conclusions and outlook}
A generator for bremsstrahlung photons and scattered electrons in high energy electron-proton and electron-ion collisions
was presented in this paper. It is ready to be used in studies of instrumentation related to luminosity measurements via
the detection of the bremsstrahlung photons, as well as studies of dedicated detectors for electrons scattered at small angles.

Its potential application extends to the determination of beam losses due to bremsstrahlung emission and as a fast generator
for final state electrons in deep inelastic scattering.

New development is foreseen to incorporate more precise physics models and the effects of a non-negligible crossing angle
of the electron and proton (ion) beams. Moreover, the generator illustrates the feasibility of using Python for event
generators in the field of high energy physics.

\section*{Acknowledgements}
The author wishes to thank
BNL colleagues for invaluable comments and suggestions.
The author acknowledges support from the U.S. Department of Energy under contract number de-sc0012704.


\bibliographystyle{elsarticle-num}
\bibliography{references}

\end{document}